%% file: cmd3_7pi.tex
\documentclass[12pt]{elsarticle}
\usepackage{refmerge}
\usepackage{epsfig}
\begin{document}
\date{\today}

\title{\bf{ \boldmath
STUDY OF THE PROCESS $e^+e^-\to 3(\pi^+\pi^-)\pi^0$
IN THE C.M. ENERGY RANGE 1.6--2.0 GEV WITH THE CMD-3 DETECTOR
}}

\input{authors_7pi_cmd3.tex}


%
\vspace{0.7cm}
\begin{abstract}
\hspace*{\parindent}
The cross section of the process $e^+e^- \to 3(\pi^+\pi^-)\pi^0$ has been 
measured for the first time using a data sample of 56.7 pb$^{-1}$ collected 
with the CMD-3 detector at the VEPP-2000  $e^+e^-$ collider.  632$\pm$32 
signal events have been selected in the center-of-mass energy range 
1.6 -- 2.0 GeV. A study of dynamics of seven-pion production allows one
to extract contributions of the dominated $2(\pi^+\pi^-)\omega$ and $2(\pi^+\pi^-)\eta$ 
intermediate states. 
\end{abstract}

\maketitle
\baselineskip=17pt
\section{ \boldmath Introduction}
\hspace*{\parindent}
Production of seven pions in $e^+e^-$ annihilation
has not been studied before. A partial estimate of the cross section is
possible from the BaBar measurement of the cross section of the 
$e^+e^- \to 2(\pi^+\pi^-)\eta, \eta\to\gamma\gamma$~\cite{isr5pi} reaction,
based on the Initial-State Radiation (ISR) method.
Using the well-known $\eta\to\pi^+\pi^-\pi^0$ decay rate, a contribution to
the seven-pion cross section can be calculated. 
As a part of the total hadronic cross section, the cross section of  
$e^+e^- \to 3(\pi^+\pi^-)\pi^0$ is interesting for   
the calculations of the hadronic contribution to the muon anomalous magnetic 
moment~\cite{g21,g22,g23}. The detailed study of the production dynamics can
further improve the accuracy of these calculations and can help  
explain energy dependence of the cross section.

In this paper we report the analysis of the data sample based on 
56.7 pb$^{-1}$ of the integrated luminosity collected at the CMD-3 detector
in the 1.6-2.0 GeV center-of-mass (c.m.) energy range. 
These data were collected in four energy scans, about 50 c.m. energy
points each, performed  at  the VEPP-2000 $e^+e^-$ 
collider~\cite{vepp1,vepp2,vepp3,vepp4} in the 
2011, 2012 and 2017 experimental runs. 
In the 2017 experimental run the beam energy has been monitored by 
the back-scattering-laser-light system~\cite{laser1,laser2}, providing
an absolute energy measurement with better than 0.1 MeV uncertainty in every single measurement. In earlier runs beam energy has been determined using charge track
momenta in detector magnetic field with about 1 MeV uncertainty.
Since the cross section of the process is small, we combine our scanned 
points into eight energy intervals as shown in Table~\ref{xs_all}.

The general-purpose detector CMD-3 has been described in 
detail elsewhere~\cite{sndcmd3}. Its tracking system consists of a 
cylindrical drift chamber (DC)~\cite{dc} and double-layer multiwire 
proportional 
Z-chamber, both also used for a trigger, and both inside a thin 
(0.2~X$_0$) superconducting solenoid with a field of 1.3~T.
The liquid xenon (LXe) barrel calorimeter with a 5.4~X$_0$ thickness has
fine electrode structure, providing 1-2 mm spatial resolution~\cite{lxe}, and
shares the cryostat vacuum volume with the superconducting solenoid.     
The barrel CsI crystal calorimeter with a thickness 
of 8.1~X$_0$ is placed
outside  the LXe calorimeter,  and the end-cap BGO calorimeter with a 
thickness of 13.4~X$_0$ is placed inside the solenoid~\cite{cal}.
The luminosity is measured using events of Bhabha scattering 
at large angles with about 1\% systematic uncertainty~\cite{lum}. 
\section{Selection of $e^+e^-\to 3(\pi^+\pi^-)\pi^0$ events}
\label{select}
\hspace*{\parindent}

The analysis procedure is based on our study of the six-charged-pion
reaction described in Ref.~\cite{cmd6pi}. 
Candidate events
are required to have six charged-particle tracks, each one having: 
\begin{itemize}
\item{
more than five hits in the DC.
}
\item{
a momentum is larger than 40 MeV/c.
}
\item{
a minimum distance from a track to the beam axis in the
transverse  plane is less than 0.5 cm.
}
\item{
a minimum distance from a track to the center of the interaction region along
the beam axis Z  is less than 10 cm.
}
\item{
a polar angle large enough to cross half of the DC radius.
}
\end{itemize} 

Reconstructed momenta and angles of the tracks for six-track events were used
for further selection. 

\begin{figure}[tbh]
\begin{center}
\vspace{-0.7cm}
\includegraphics[width=1.0\textwidth]{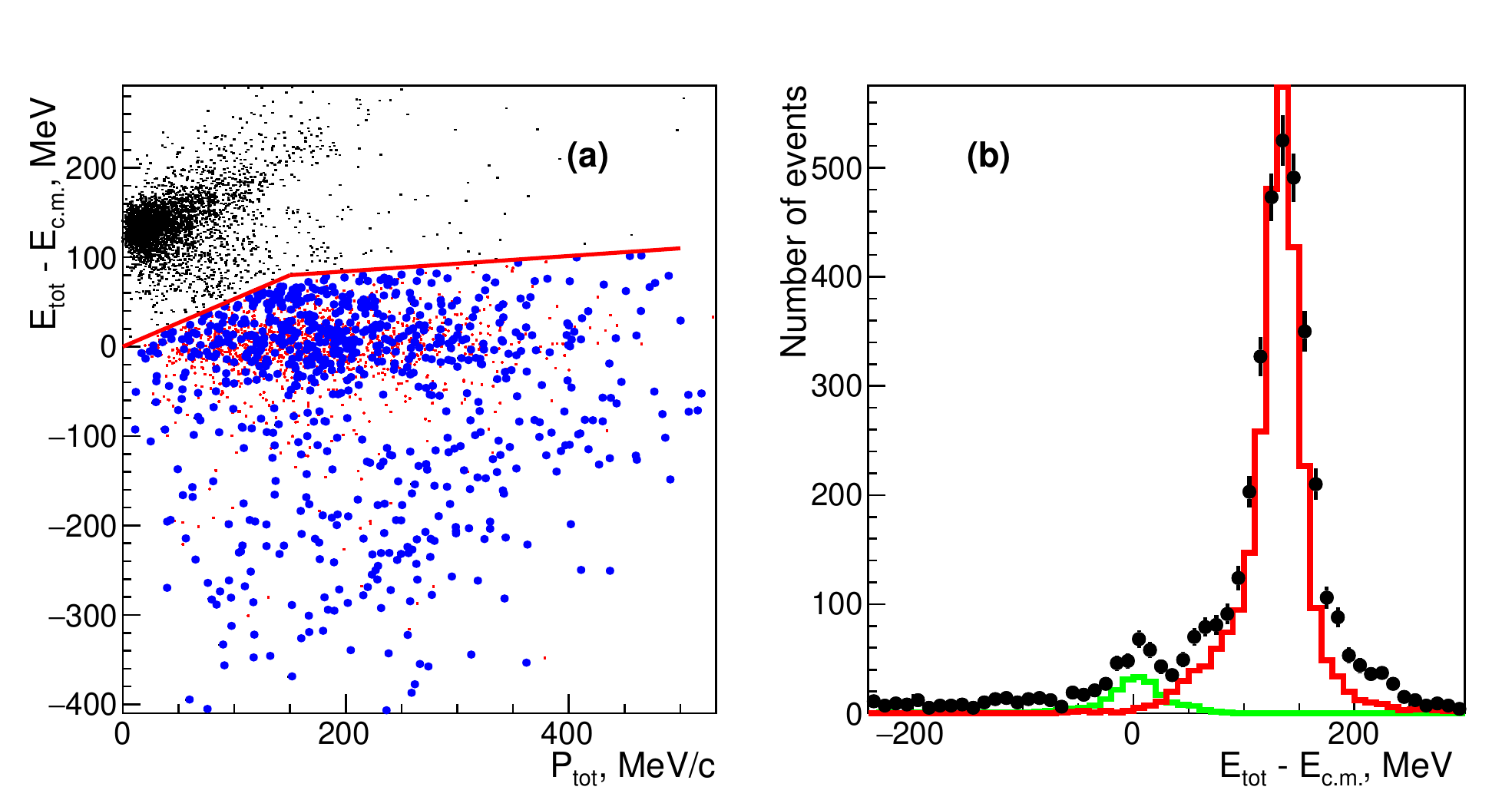}
\vspace{-0.3cm}
\caption
{
(a) Scatter plot of the difference  between the energy of seven pions 
and c.m. energy ($\Delta E$) vs total momentum. The line shows 
the boundary of the applied selection, where data points are shown by 
increased circles, and seven-pion signal simulation is shown by red 
(in color version) crosses; 
(b) Projection plot of (a). 
The solid histograms show the normalized MC-simulated distribution for the  
expected seven-pion signal (left peak) and six-pion background (right peak).
}
\label{6energy}
\end{center}
\end{figure}

The analysis strategy is based on the reconstruction of the six-charged-pion
system, assuming a missing $\pi^0$ particle. The total energy $\rm E_{tot}$ of the seven
pion final state is calculated from the total momentum $\rm P_{tot}$ of charged tracks:

$$
\rm P_{tot} = \large |\sum_{i=1}^{6}\bar p_{i}\large |, 
~~~\rm E_{tot} = \sum_{i=1}^{6}\sqrt{p_{i}^2+m_{\pi}^2} + \sqrt{P_{tot}^2+m_{\pi^0}^2}~.
$$

We do not use calorimeter
responce for the photons from the $\pi^0$ decay due to large number
of extra soft clusters from the charge pion nuclear interactions. These clusters are not properly reproduced
in simulation.

Figure~\ref{6energy}(a) shows a scatter plot of the difference between 
the total energy and c.m. energy, $\Delta \rm E = E_{tot}-E_{c.m.}$, vs the total
momentum $\rm P_{tot}$ for the six-track candidates. 
A clear signal of the $e^+ e^- \to 3(\pi^+\pi^-)$ reaction is seen in data as a 
cluster of dots 
at $\Delta \rm E = 135$ MeV and the total momentum near zero.
The expected seven-pion signal has the $\Delta \rm E $ value near zero, 
and the $\rm P_{tot}$ value is distributed up to 400 MeV/c, as shown by 
the (red) crosses from the Monte Carlo (MC) signal simulation. The enlarged 
(blue) circles show data in the region where we search for signal events.
Figure~\ref{6energy}(b) shows the projection plot of (a): circles are for 
the data and the histograms show the normalized to data MC-simulated distributions 
for the seven-pion signal and six-pion background.

\begin{figure}[tbh]
\begin{center}
\vspace{-0.2cm}
\includegraphics[width=1.0\textwidth]{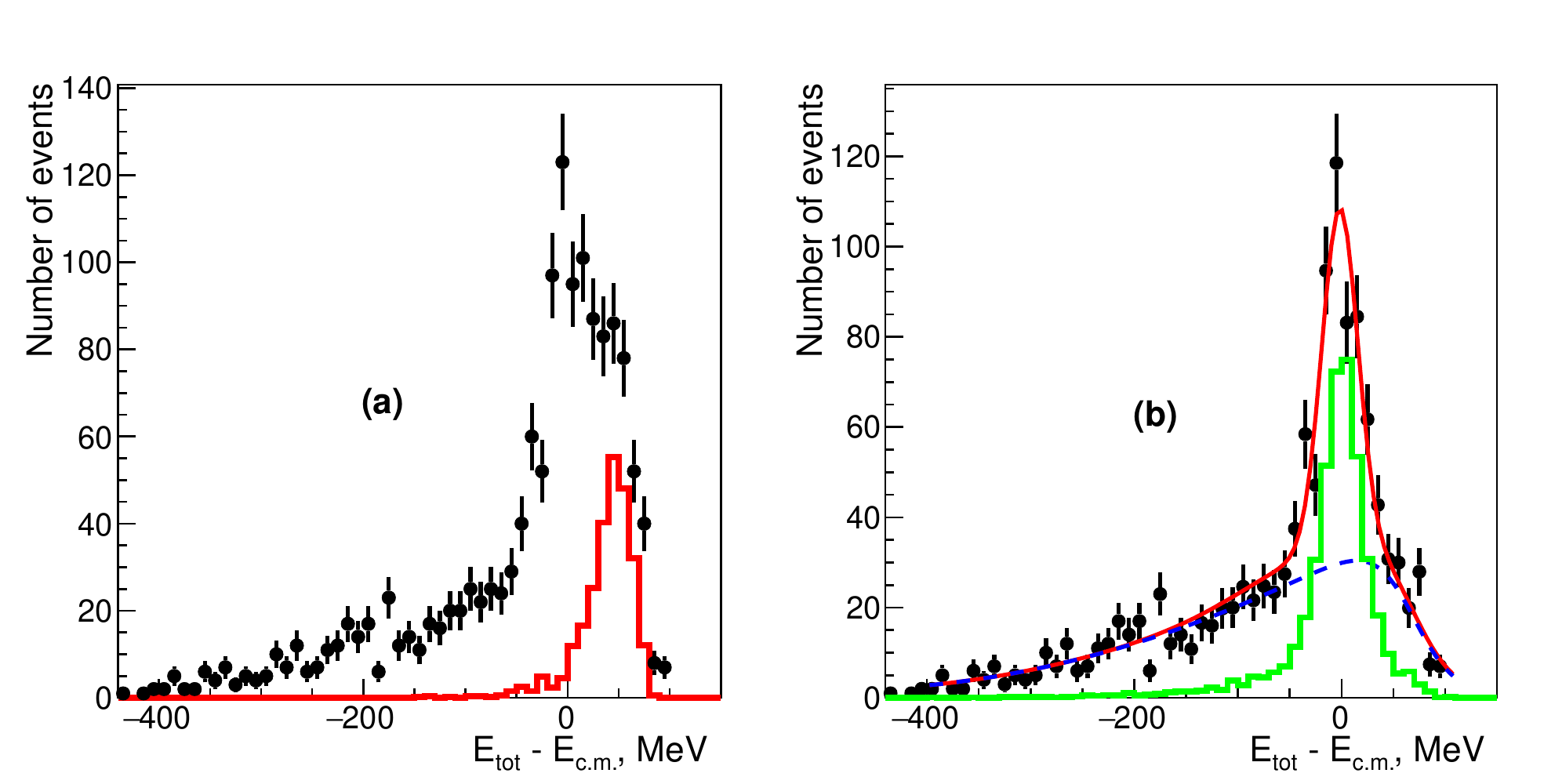}
\vspace{-0.3cm}
\caption
{
(a) The difference  between the energy of seven pions and c.m. energy ($\Delta E$) after selection by the line in Fig.~\ref{6energy}(a). All energy intervals are summed. The histogram shows  
the normalized to the data of Fig.~\ref{6energy}(b) MC-simulated distribution for the remaining six-pion background.
(b) Example of the fit to the seven-pion signal (solid line) and remaining background (dashed line) after the six-pion background subtraction.  The histogram shows the expected signal from the MC simulation.
}
\label{EnergyCut}
\end{center}
\end{figure}

To reduce a contribution from six-pion events, we select events below the 
line shown in Fig.~\ref{6energy}(a). The $\Delta E$ distribution of the event candidates after selection is shown in Fig.~\ref{EnergyCut}(a) by circles, while the histogram shows the remaining contribution of the six-pion events. 
The observed signal of six-pion events at each energy interval is used to normalize the MC simulation.
We subtract this 
contribution from the experimental distribution of Fig.~\ref{EnergyCut}(a), 
and show the result in Fig.~\ref{EnergyCut}(b) together with the fit functions 
used to determine the number of seven-pion events and remaining background. 
The signal line shape is taken from the MC simulation 
of the seven-pion process, shown by the histogram, and is well described by 
the double-Gaussian function. All parameters of the signal function are fixed
according to MC simulation at each energy interval  
except for the number of events and the main Gaussian resolution.
A third-order polynomial is used to describe the remaining background 
distribution shown by the dashed line in  Fig.~\ref{EnergyCut}(b).

A variation of the polynomial  parameters for the experimental and 
MC-simulated signal distributions as well as variation of applied selections
lead to an about 10\% uncertainty on the number of signal events, which is taken as an estimate of the systematic uncertainty. The background contribution increases with energy, and for the highest energy interval we estimate this uncertainty as 15\%.

We apply this procedure to the event sample in each energy interval, and 
in total find 632$\pm$32 signal events, corresponding to the process 
$e^+e^-\to 3(\pi^+\pi^-)\pi^0$ in the studied energy range. 
The numbers of selected events determined in each energy interval are 
listed in Table~\ref{xs_all}.
\section{First study of the production dynamics}
\label{dynamics}
\begin{center}
\begin{figure}[tbh]
\vspace{-0.2cm}
\includegraphics[width=0.5\textwidth]{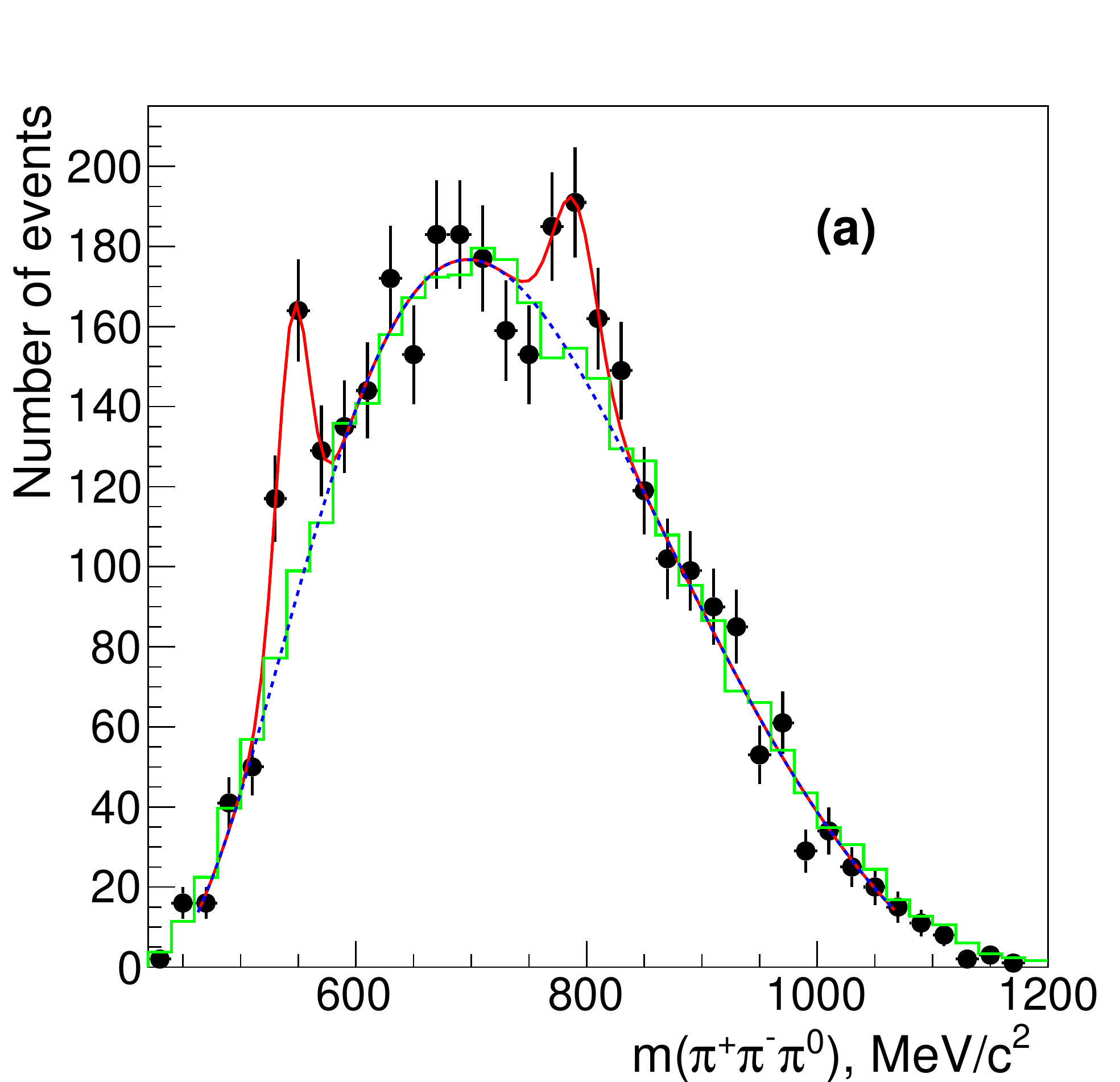}
\includegraphics[width=0.54\textwidth]{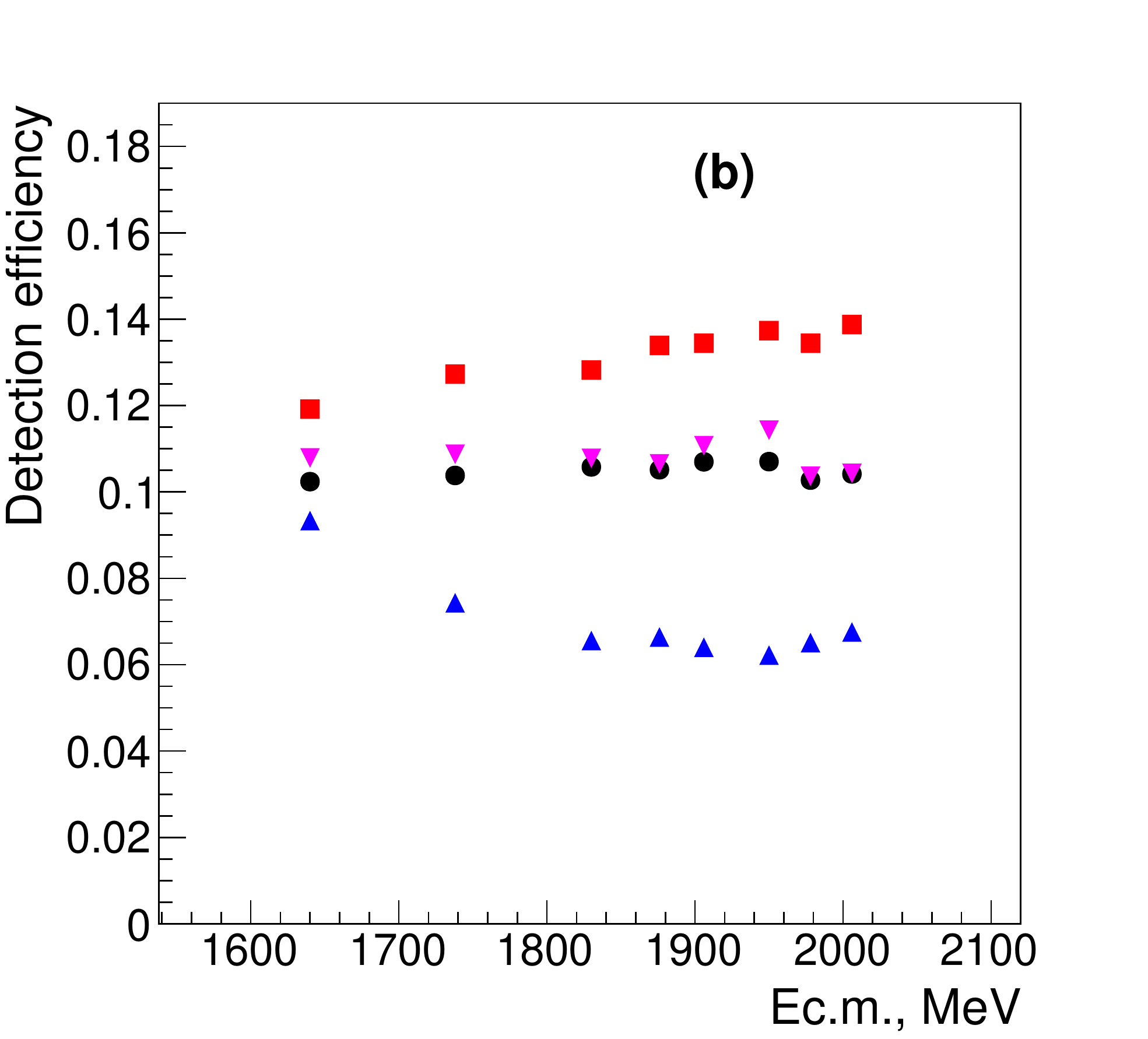}
\caption
{(a) Experimental $\pi^+\pi^-\pi^0$ invariant mass distribution (nine entries per event) for the events from the signal peak of Fig.~\ref{EnergyCut}(b). The 
solid line shows the fit functions describing the signals from 
$\eta$, $\omega$, and the combinatorial background (dashed curve). The 
histogram represents MC simulation in the phase-space model. 
(b) Detection efficiency obtained from the MC simulation for  the $2(\pi^+\pi^-)\omega$ model (squares), and for the $2(\pi^+\pi^-)\eta$ one (circles) in 
case of extracting events from the $\Delta \rm E$ peak of Fig.~\ref{EnergyCut}(b), or from the $\eta$ and the $\omega$ signals in the three-pion mass distribution  (triangles and up-down triangles, respectively).  
}
\label{masses}
\end{figure}
\end{center}

\hspace*{\parindent}
The dynamics of the process $e^+e^-\to 3(\pi^+\pi^-)\pi^0$   
has not been studied previously. The BaBar Collaboration~\cite{isr5pi}
reported the observation of the $e^+e^-\to 2(\pi^+\pi^-)\eta, ~\eta\to\gamma\gamma$ process, which contributes to seven final-state pions if the 
$\eta$ decays to $\pi^+\pi^-\pi^0$.
We investigate the production mechanisms using the events in the signal region 
of Fig.~\ref{EnergyCut}(b) using the requirement $|\Delta \rm E| <$ 60 MeV. 
Figure~\ref{masses}(a) shows an invariant mass distribution for all 
$\pi^+\pi^-\pi^0$ combinations (nine entries per event) for selected events. 
The signal from the $\eta$ meson is clearly seen, as well as presence of the 
$\omega(782)$ resonance in the intermediate state with the $\omega\to\pi^+\pi^-\pi^0$  decay. To obtain the number of events with $\eta$ and $\omega$ in the 
intermediate states, we fit this distribution with the sum of functions 
describing combinatorial background and the peaks from the $\eta$ and the 
$\omega$ signals as shown by the solid curve in Fig.~\ref{masses}(a).  
Our resolution is significantly larger than the resonance widths 
(about 20 MeV), and we use the  Gaussian function for the peaks, while the 
polynomial function is used for the combinatorial background (the dashed line 
in Fig.~\ref{masses}(a)). The combinatorial background is well described by 
the MC-simulated distribution in the phase-space model without any intermediate 
resonances, shown by the histogram in Fig.~\ref{masses}(a). In total, we 
obtain $280\pm36$ events for the  $e^+e^-\to 2(\pi^+\pi^-)\eta,~\eta\to\pi^+\pi^-\pi^0$  process and $204\pm37$ events for $e^+e^-\to 2(\pi^+\pi^-)\omega,~\omega\to\pi^+\pi^-\pi^0$. Note that the total number of the $3(\pi^+\pi^-)\pi^0$ 
events ($632\pm32$) exceeds the sum of the events from the $\eta$ and $\omega$ peaks  ($484\pm52$) by about 32\%: this is discussed below.

We apply this fit to every energy interval and list the obtained number of 
events in Table~\ref{xs_all}.

We calculate the invariant masses for the combinations of 
the two (total charge $\pm 1$ or zero), the four (total charge zero), and the 
five (total charge zero) pions from the selected events, and find no signal 
from the $\rho(770)$ resonance or from any other resonances in our range of 
the c.m. energies. In general, all these distributions are well described 
by the phase-space model.
\section{Detection efficiency}
\label{efficiency}
\hspace*{\parindent}

In our experiment, the acceptance of the DC for charged tracks is not 
100\%, and
the detection efficiency depends on the production dynamics of seven pions.
We have developed the primary generators for the seven-pion final-state 
production in the $e^+e^-$ collision for the phase-space model, and for the 
models with the intermediate $2(\pi^+\pi^-)\eta$  and  $2(\pi^+\pi^-)\omega$ 
states. In our model the $2(\pi^+\pi^-)\eta$ intermediate state is described 
as the  $\rho(1450)\eta$ production with the $\rho(1450)$ decay either to four 
pions in the P-wave or to the $a_1(1260)\pi$ state. The $2(\pi^+\pi^-)\omega$ 
state is modeled as production of the $f_0(1370)\omega$ state, followed by 
four pions from the $f_0(1370)$ decay in the S-wave.  

To obtain the detection efficiency, we simulate seven-pion production 
in the primary generators, pass simulated events through the 
CMD-3 detector using the GEANT4~\cite{geant4} package, and reconstruct them
with the same reconstruction software as experimental data.  
We calculate the detection efficiency from the MC-simulated events
as a ratio of events after the selections described in 
Secs.~\ref{select},\ref{dynamics} 
to the total number of generated events. 

Figure~\ref{masses}(b) shows the detection efficiency obtained for 
the $2(\pi^+\pi^-)\omega$ (squares) and for the $2(\pi^+\pi^-)\eta$ (circles) 
intermediate states  when the number of signal events is obtained from the fit
of the $\Delta \rm E$ peak of Fig.~\ref{EnergyCut}(b).  Due to the difference 
in the angular and momentum distributions of the pions, the efficiency for 
the $2(\pi^+\pi^-)\eta$ intermediate state is lower compared to the 
$2(\pi^+\pi^-)\omega$ model: about 10\% and 13\%, respectively. Variations of 
the dynamics or resonance parameters inside the initial 
``vector-pseudo-scalar'' state for the  $2(\pi^+\pi^-)\eta$ production, 
and inside the ``scalar-vector'' state for the $2(\pi^+\pi^-)\omega$ 
production do not change the obtained detection efficiency by more than 3--5\%. 

If we determine the number of the MC-simulated events using the $\eta$ and 
$\omega$ peaks from the histogram similarly to that in Fig.~\ref{masses}(a), 
the detection efficiency decreases 
additionally by 20--40\% due to the $|\Delta \rm E| <$ 60 MeV requirement. 
These efficiencies are shown in Fig.~\ref{masses}(b) by triangles and up-down 
triangles for the $2(\pi^+\pi^-)\eta$  and  $2(\pi^+\pi^-)\omega$ states, 
respectively.
\section{Cross section calculation}    
\hspace*{\parindent}
In each energy interval the cross section is calculated as 
$$
\sigma = \frac{N}{L\cdot\epsilon\cdot(1+\delta)},
$$ 
where $N$ is the number of signal events,
$L$ is the integrated luminosity for this energy interval, $\epsilon$ is 
the detection efficiency, 
and $(1+\delta)$ is the radiative correction calculated 
according to Ref.~\cite{kur_fad,actis}.
To calculate the radiative correction, we use BaBar data for the $e^+e^-\to 2(\pi^+\pi^-)\eta$ reaction ~\cite{isr5pi} as a first 
approximation, and obtain $(1+\delta) = 0.92$ with very weak energy 
dependence.  

\begin{center}
\begin{figure}[tbh]
\vspace{-0.2cm}
\includegraphics[width=0.5\textwidth]{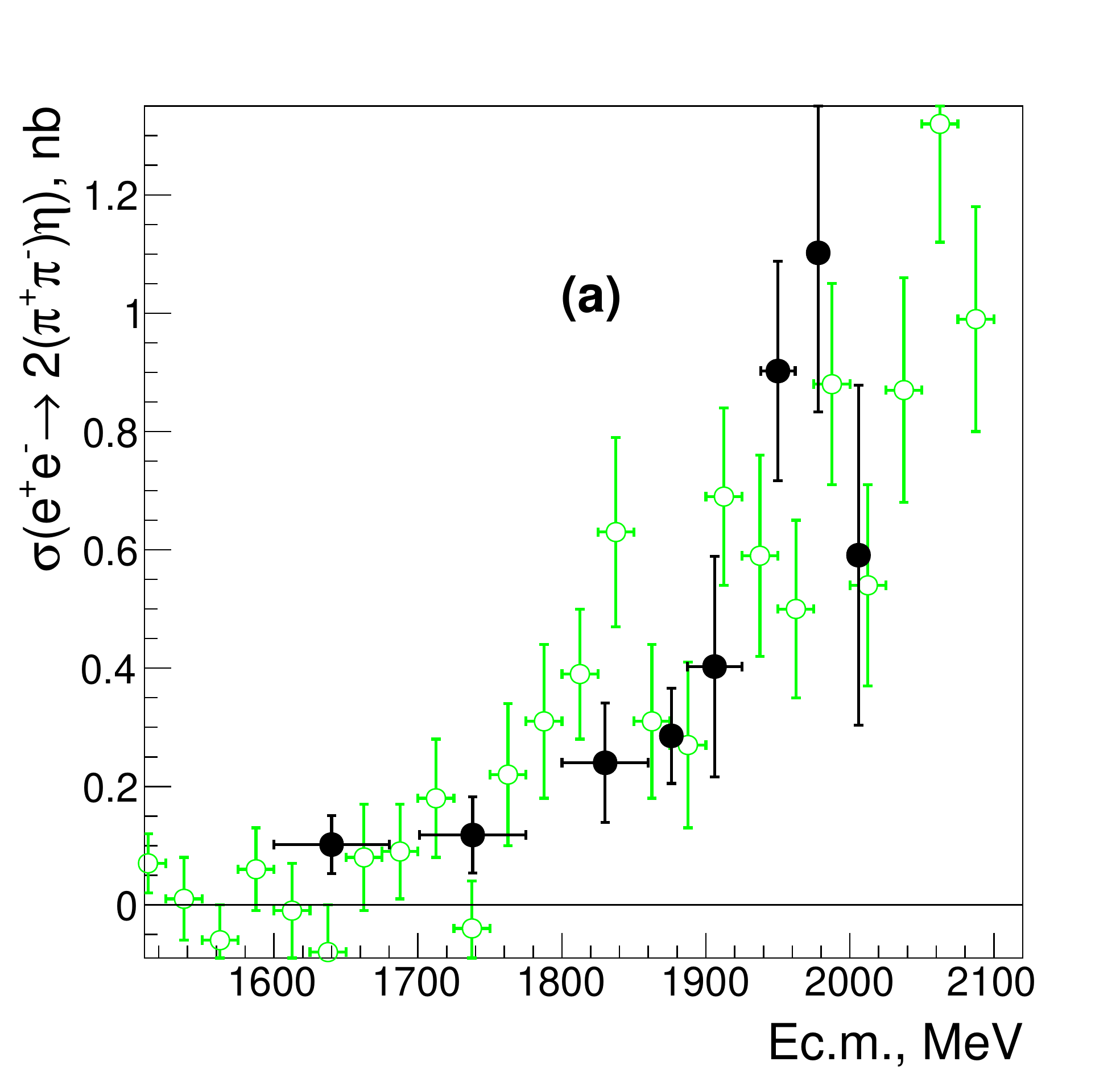}
\includegraphics[width=0.5\textwidth]{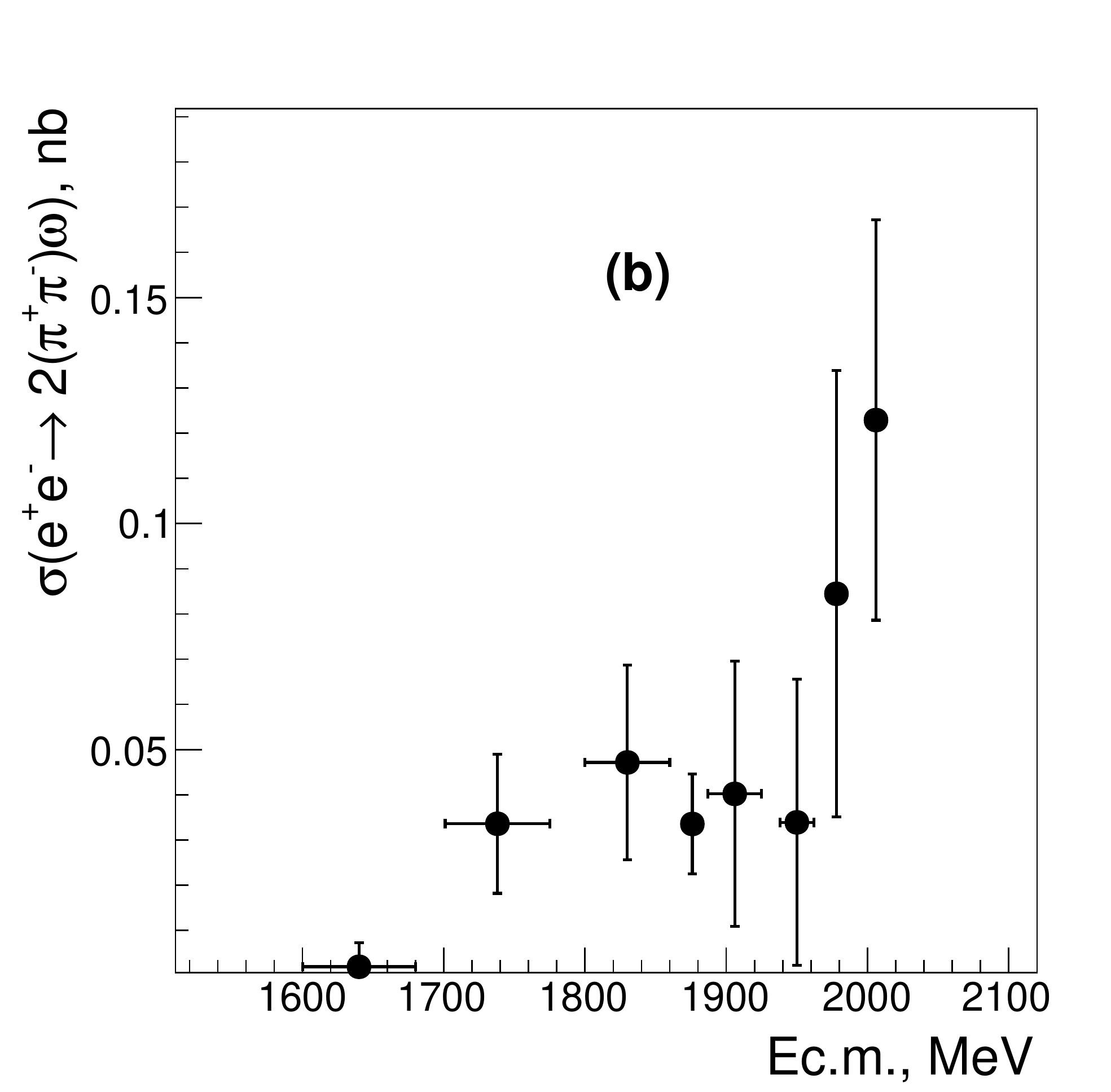}
\caption{
(a) The $e^+e^-\to 2(\pi^+\pi^-)\eta$ cross section measured with the CMD-3
detector at VEPP-2000 (circles). The results of the BaBar
measurement~\cite{isr5pi} are shown by open circles. 
(b) The $e^+e^-\to 2(\pi^+\pi^-)\omega$ cross section measured with the CMD-3
detector at VEPP-2000.
}
\label{cross}
\end{figure}
\end{center}

We calculate the cross sections for the $e^+e^-\to 2(\pi^+\pi^-)\eta$ and $e^+e^-\to 2(\pi^+\pi^-)\omega$ reactions using the efficiencies shown by triangles 
and up-down triangles in Fig.~\ref{masses}(b), respectively. These cross 
sections are shown in Fig.~\ref{cross}(a,b): the branching fractions of the $\eta\to\pi^+\pi^-\pi^0$ and $\omega\to\pi^+\pi^-\pi^0$ decays are taken into account using values from Ref.~\cite{pdg}. We observe relatively good agreement with the BaBar measurement of the $e^+e^-\to 2(\pi^+\pi^-)\eta$ reaction, while no other measurements exist for the $e^+e^-\to 2(\pi^+\pi^-)\omega$ cross section. 

As mentioned in Secs.~\ref{select},\ref{dynamics}, the total number of  
$3(\pi^+\pi^-)\pi^0$ events is about 32\% larger than the sum of the 
individual channels with the $\eta$ and $\omega$ intermediate states.  
This difference is almost eliminated after taking into account the  
difference in the efficiency obtained by the fit of $\Delta \rm E$ or by 
the fit of the $\eta$ and $\omega$ signals where cut $|\Delta \rm E < 60|$ MeV is applied: the average ratios are about 1.35--1.37 for both channels. The obtained number 
$\rm N_{eff} =((280\pm36 ) + (204\pm37))\cdot 1.36 = 658\pm70$ is  
consistent with the total number of the $3(\pi^+\pi^-)\pi^0$ events ($632\pm32$) within the statistical uncertainty. We come to the conclusion that the 
inclusive $e^+e^-\to 3(\pi^+\pi^-)\pi^0$ cross section is completely dominated 
by the sum of the two intermediate states within the measured accuracy. 

\begin{center}
\begin{figure}[tbh]
\vspace{-0.2cm}
\includegraphics[width=1.0\textwidth]{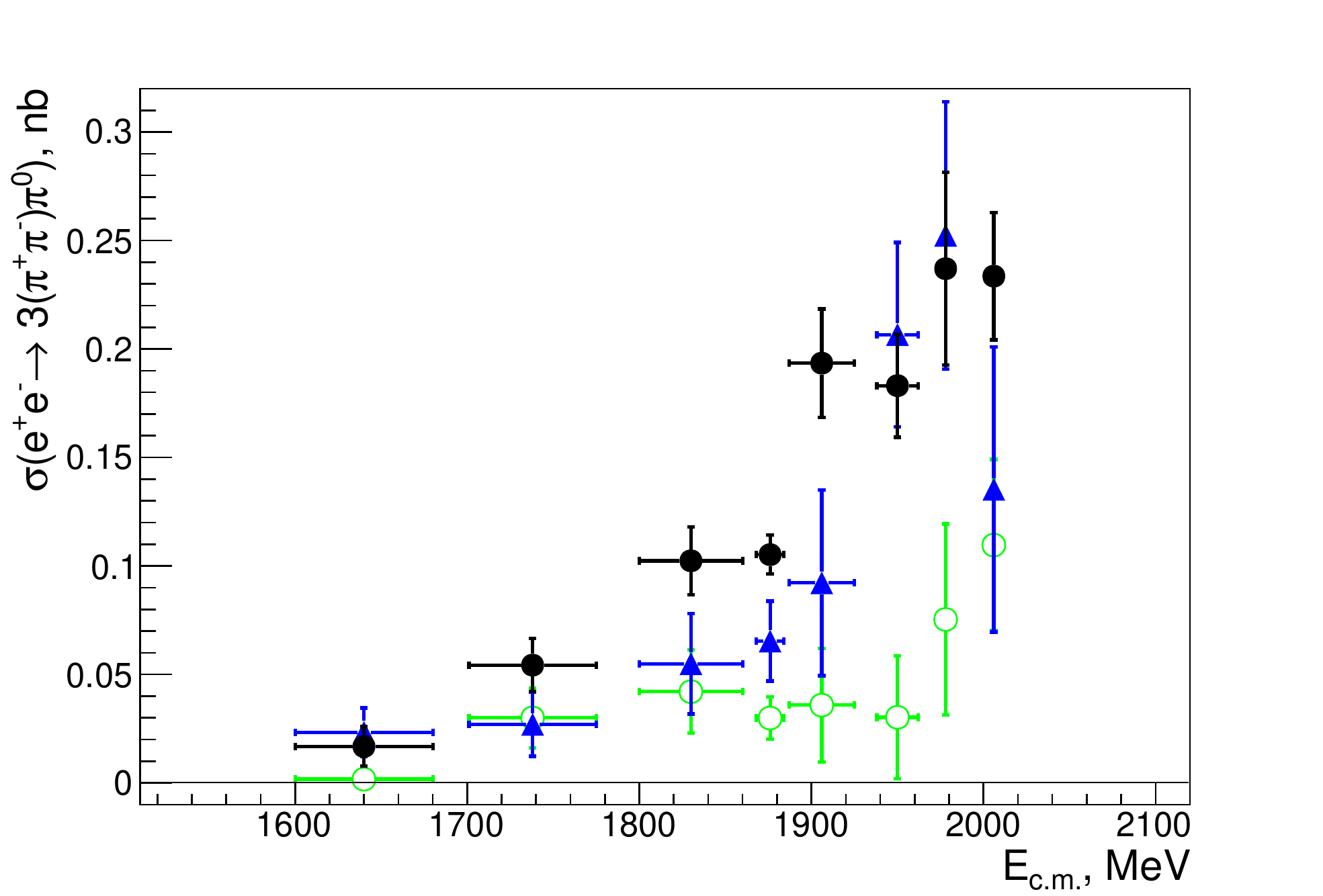}
\caption{
The $e^+e^-\to 3(\pi^+\pi^-)\pi^0$ cross section measured with the CMD-3
detector at VEPP-2000 (dots). The contribution from the $e^+e^-\to 2(\pi^+\pi^-)\eta$ and $e^+e^-\to 2(\pi^+\pi^-)\omega$ reactions are shown by triangles and open circles, respectively.
}
\label{crossall}
\end{figure}
\end{center}

To calculate the inclusive cross section for the $e^+e^-\to 3(\pi^+\pi^-)\pi^0$ reaction,  we average the efficiencies in Fig.~\ref{masses}(b) for the $\eta$ and $\omega$ intermediate states with the weight corresponding to the ratio 
of the corrected number of the events: the $2(\pi^+\pi^-)\eta$ efficiency 
is taken with the 1.18 weight value. The resulting cross section is shown 
in Fig.~\ref{crossall} by circles. We assign an additional 15\% uncertainty 
due to statistical fluctuations of the ratio.

For comparison, we show in Fig.~\ref{crossall} the contribution from the 
$e^+e^-\to 2(\pi^+\pi^-)\eta$ and $e^+e^-\to 2(\pi^+\pi^-)\omega$ reactions by 
triangles and open circles, respectively: only decays of $\eta$ and $\omega$ 
to three pions are taken. 
The $e^+e^-\to\pi^+\pi^-\eta'(958)$  reaction, reported in Ref.~\cite{isr5pi}, contributes about 0.1 nb to the total cross section at $E_{c.m.} = 2.0$ GeV, but the decay rate of $\eta'(958)\to 2(\pi^+\pi^-)\eta\to 3(\pi^+\pi^-)\pi^0$ to the studied final state reduces the visible cross section to 0.01 nb, what is less than a sensitivity of our experiment.

The integrated luminosity, the number of the seven-pion events, the number 
of events for the $2(\pi^+\pi^-)\eta$  and  $2(\pi^+\pi^-)\omega$ intermediate 
states, and obtained cross sections for each energy interval are listed in 
Table~\ref{xs_all}. 


\section{Systematic uncertainties}
\hspace*{\parindent}
The following sources of systematic uncertainties are considered.

\begin{itemize}

\item{The tracking efficiency was studied in detail in our previous 
papers~\cite{cmd4pi,cmd6pi}, and the correction for the track reconstruction 
efficiency compared to the MC simulation is about 1.5$\pm$1.0\% per track: 
the MC-simulated detection efficiency is corrected by -6\% while 3\% is taken 
as the corresponding systematic uncertainty. 
}
\item {
The model dependence of the acceptance is determined using 
the comparison of efficiencies calculated for the different production 
dynamics for $e^+e^-\to 2(\pi^+\pi^-)\eta$ and the $e^+e^-\to 2(\pi^+\pi^-)\omega$ reactions. It is estimated as 3-5\%.
}  
\item{
Since only one charged track is 
sufficient for a trigger (98-99\% efficiency), we assume 
that for the multitrack events considered in this analysis the 
trigger inefficiency gives a negligible contribution to the systematic uncertainty. 
}
\item{
A systematic uncertainty due to the selection criteria is studied by 
varying the requirements described above and doesn't exceed 5\%. 
}
\item{
The uncertainty on the determination of the integrated luminosity 
comes from the selection criteria of Bhabha events, radiative
corrections and calibrations of DC and CsI and does not exceed 
1\%~\cite{lum}.
}
\item{
The uncertainty in the background subtraction is studied
by the variation of the functions used for the background description in the 
fit, shown in Fig.~\ref{EnergyCut}(b) and is estimated as 10\% (15\% for 
$E_{c.m.} = 2.0$ GeV). 
}
\item{
The radiative correction uncertainty is estimated as about 
2\%, mainly due to the uncertainty on the maximum allowed energy of the 
emitted photon, as well as from the uncertainty on the cross section.
}
\item{
For the inclusive $e^+ e^-\to 3(\pi^+\pi^-)\pi^0$ cross section we introduce 
an additional 15\% systematic uncertainty due to the difference in the 
efficiency for the $2(\pi^+\pi^-)\eta$ and $2(\pi^+\pi^-)\omega$  
intermediate states.
}
\end{itemize}

The above systematic uncertainties summed in quadrature give an overall
systematic error of about 13\%, increasing to 20\% for the inclusive 
cross section. 

\input{table_7pi.tex}

\section*{ \boldmath Conclusion}
\hspace*{\parindent}
The total cross section of the process $e^+e^-\to 3(\pi^+\pi^-)\pi^0$ 
has been measured for the first time using 56.7 pb$^{-1}$ of integrated 
luminosity 
collected by the CMD-3 detector at the VEPP-2000 $e^+e^-$ collider
in the 1.6-2.0 GeV c.m. energy range. 
From our study  we can conclude that the observed cross section can be 
described by the 
$e^+e^-\to 2(\pi^+\pi^-)\eta$ and the $e^+e^-\to 2(\pi^+\pi^-)\omega$ reactions.
The measured cross section for the $e^+e^-\to 2(\pi^+\pi^-)\eta$ reaction is 
in good agreement with the only available measurement by BaBar~\cite{isr5pi}.

\subsection*{Acknowledgements}
\hspace*{\parindent}
The authors are grateful to A.I.~Milstein
for his help with theoretical interpretation and development of
the models. 
We thank the VEPP-2000 team for excellent machine operation. 
The work is partially supported by the Russian 
Foundation for Basic Research grants 18-32-01020, 17-52-50064.

\input{biblio_7pi_cmd3.tex}
\end{document}

%% file: authors_7pi_cmd3.tex
\author[adr1,adr2]{R.R.~Akhmetshin}
\author[adr1,adr2]{A.N.~Amirkhanov}
\author[adr1,adr2]{A.V.~Anisenkov}
\author[adr1,adr2]{V.M.~Aulchenko}
\author[adr1]{V.Sh.~Banzarov}
\author[adr1]{N.S.~Bashtovoy}
\author[adr1,adr2]{D.E.~Berkaev}
\author[adr1,adr2]{A.E.~Bondar}
\author[adr1]{A.V.~Bragin}
\author[adr1,adr2,adr5]{S.I.~Eidelman}
\author[adr1,adr2]{D.A.~Epifanov}
\author[adr1,adr2,adr3]{L.B.~Epshteyn}
\author[adr1,adr2]{A.L.~Erofeev}
\author[adr1,adr2]{G.V.~Fedotovich}
\author[adr1,adr2]{S.E.~Gayazov}
\author[adr1,adr2]{A.A.~Grebenuk}
\author[adr1,adr2]{S.S.~Gribanov}
\author[adr1,adr2,adr3]{D.N.~Grigoriev}
\author[adr1,adr2]{F.V.~Ignatov}
\author[adr1,adr2]{V.L.~Ivanov}
\author[adr1]{S.V.~Karpov}
\author[adr1,adr2]{V.F.~Kazanin}
\author[adr1,adr2]{I.A.~Koop}
\author[adr1]{A.N.~Kirpotin}
\author[adr1,adr2]{A.A.~Korobov}
\author[adr1,adr3]{A.N.~Kozyrev}
\author[adr1,adr2]{E.A.~Kozyrev}
\author[adr1,adr2]{P.P.~Krokovny}
\author[adr1,adr2]{A.E.~Kuzmenko}
\author[adr1,adr2]{A.S.~Kuzmin}
\author[adr1,adr2]{I.B.~Logashenko}
\author[adr1,adr2]{P.A.~Lukin}
\author[adr1]{K.Yu.~Mikhailov}
\author[adr1]{V.S.~Okhapkin}
\author[adr1]{A.V.~Otboev}
\author[adr1]{Yu.N.~Pestov}
\author[adr1,adr2]{A.S.~Popov}
\author[adr1,adr2]{G.P.~Razuvaev}
\author[adr1]{Yu.A.~Rogovsky}
\author[adr1]{A.A.~Ruban}
\author[adr1]{N.M.~Ryskulov}
\author[adr1,adr2]{A.E.~Ryzhenenkov}
\author[adr1]{A.I.~Senchenko}
\author[adr1]{Yu.M.~Shatunov}
\author[adr1]{P.Yu.~Shatunov}
\author[adr1,adr2]{V.E.~Shebalin}
\author[adr1,adr2]{D.N.~Shemyakin}
\author[adr1,adr2]{B.A.~Shwartz}
\author[adr1,adr2]{D.B.~Shwartz}
\author[adr1,adr4]{A.L.~Sibidanov}
\author[adr1,adr2]{E.P.~Solodov\fnref{tnot}}
\author[adr1]{V.M.~Titov}
\author[adr1,adr2]{A.A.~Talyshev}
\author[adr1]{A.I.~Vorobiov}
\author[adr1]{I.M.~Zemlyansky}
\author[adr1,adr2]{Yu.V.~Yudin}

\address[adr1]{Budker Institute of Nuclear Physics, SB RAS, 
Novosibirsk, 630090, Russia}
\address[adr2]{Novosibirsk State University, Novosibirsk, 630090, Russia}
\address[adr3]{Novosibirsk State Technical University, 
Novosibirsk, 630092, Russia}
\address[adr4]{University of Victoria, Victoria, BC, Canada V8W 3P6}
\address[adr5]{Lebedev Physical Institute RAS, Moscow, 119333, Russia}
\fntext[tnot]{Corresponding author: solodov@inp.nsk.su}

%% file: table_7pi.tex
\begin{table}[tbh]
\caption{Energy interval, integrated luminosity, number of signal seven-pion events  
and obtained cross sections for the
$e^+e^-\to 3(\pi^+\pi^-)\pi^0$, $e^+e^-\to 2(\pi^+\pi^-)\eta$, and $e^+e^-\to 2(\pi^+\pi^-)\omega$ reactions.
Only statistical uncertainties are shown. 
}
\label{xs_all}
\vspace{-0.7cm}
\begin{center}
\renewcommand{\arraystretch}{0.85}
\begin{tabular}{cccccccc}
\hline
{E$_{\rm c.m.}$, MeV} & {L, nb$^{-1}$} 
&{$N_{6\pi\pi^0}$}
&{$N_{4\pi\eta}$}
&{$N_{4\pi\omega}$}
&{$\sigma_{6\pi\pi^0}$, nb}
&{$\sigma_{4\pi\eta}$, nb}
&{$\sigma_{4\pi\omega}$, nb}\\
\hline
2007.0$\pm$0.5& 4259& 95$\pm$12 &36$\pm$17& 45$\pm$16 &  0.23$\pm$0.03 & 0.59$\pm$0.29 &  0.12$\pm$0.04 \\ 
1980$\pm$1    & 2368& 53$\pm$10 &36$\pm$9 & 17$\pm$10 &  0.24$\pm$0.04 & 1.10$\pm$0.27 &  0.08$\pm$0.05\\ 
1940--1962    & 4631& 84$\pm$11 &55$\pm$11& 15$\pm$14 &  0.18$\pm$0.02 & 0.90$\pm$0.19 & 0.033$\pm$0.032\\ 
1890--1925    & 5497&105$\pm$14 &30$\pm$14& 20$\pm$15 &  0.19$\pm$0.02 & 0.40$\pm$0.19 & 0.040$\pm$0.029\\ 
1870--1884    &16803&171$\pm$15 &67$\pm$19& 49$\pm$16 & 0.105$\pm$0.009& 0.29$\pm$0.08 & 0.033$\pm$0.011\\ 
1800--1860    & 8287& 83$\pm$13 &27$\pm$12& 35$\pm$16 & 0.102$\pm$0.016& 0.24$\pm$0.10 & 0.047$\pm$0.021\\ 
1700--1775    & 7589& 39$\pm$9  &14$\pm$8 & 23$\pm$10 & 0.054$\pm$0.012& 0.11$\pm$0.06 & 0.033$\pm$0.015\\ 
1600--1680    & 7299& 12$\pm$6  &15$\pm$7 &  1$\pm$3  & 0.017$\pm$0.008& 0.10$\pm$0.05 & 0.002$\pm$0.005\\ 
\hline
\end{tabular}
\end{center}
\end{table}